\documentstyle[12pt]{article}

\newcommand {\beq}{\begin{equation}}
\newcommand {\eeq}{\end{equation}}
\newcommand {\beqa}{\begin{eqnarray}}
\newcommand {\eeqa}{\end{eqnarray}}
\newcommand {\beqan}{\begin{eqnarray*}}
\newcommand {\eeqan}{\end{eqnarray*}}
\newcommand {\n}{\nonumber \\}

\newcommand {\Romannumeral}[1]{\uppercase\expandafter{\romannumeral#1}}

\begin{document}
\setlength{\oddsidemargin}{0cm}
\setlength{\baselineskip}{7mm}

\begin{titlepage}
 \renewcommand{\thefootnote}{\fnsymbol{footnote}}
$\mbox{ }$
%\vspace{-3cm}
\begin{flushright}
\begin{tabular}{l}
OU-HET 281 \\
%{\bf KEK preprint 96 }\\
%TIT/HEP-357\\   
%hep-th/9612115\\
November 1997
\end{tabular}
\end{flushright}

%    \begin{normalsize}
%     \begin{flushright}
%                 KEK-TH-503  \\
%                     \\
%                 December 1996
~~\\
~~\\
~~\\
%~~
%     \end{flushright}

\vspace*{0cm}
    \begin{Large}
%       \vspace{1cm}
       \begin{center}
         {\Large  Exact Results in $N_c=2$ IIB Matrix Model}      \\
       \end{center}
    \end{Large}

  \vspace{1cm}

\begin{center}
           Takao S{\sc uyama}\footnote
           {e-mail address : suyama@funpth.phys.sci.osaka-u.ac.jp}
{\sc and}
           Asato T{\sc suchiya}\footnote
           {e-mail address : tsuchiya@funpth.phys.sci.osaka-u.ac.jp}\\
      \vspace{1cm}
        {\it Department of Physics, Graduate School of Science,
                Osaka University,}\\
               {\it Toyonaka, Osaka 560, Japan} \\
%      \vspace{1cm}
\end{center}

\vfill

\begin{abstract}
\noindent
We investigate $N_c=2$ case of IIB matrix model, which is exactly soluble.
We calculate the partition function exactly
%in taking care of zero modes 
and obtain a finite result without introducing any cut-off.
We evaluate some correlation functions consisting of Wilson loops.

\end{abstract}
\vfill
\end{titlepage}
\vfil\eject

%\vspace{1cm}

%\section{Introduction}
%\setcounter{equation}{0}
%\renewcommand{\thefootnote}{\fnsymbol{footnote}}
%\hspace*{\parindent}
Recently IIB matrix model has been proposed as a candidate for nonperturbative 
definitions of superstring theory \cite{IKKT}. This model is the large-N reduced
model of ten-dimensional super Yang-Mills theory. 
%It is called IIB matrix model. 
The model has manifest $N=2$ super Poincare
invariance and is related to the Schild-type action \cite{Schild} 
for the Green-Schwarz type IIB superstring \cite{GS}. 
At one-loop level it describes low-energy behavior for the interactions
of D-branes consistently.  A remarkable feature is that in this model
space-time is generated dynamically as eigenvalues or diagonal elements
of matrices.
Furthermore it has been shown in ref. \cite{FKKT} 
that the light-cone string field 
theory for type IIB superstring \cite{GSB} can be derived from 
the Schwinger-Dyson
equations for Wilson loops. This means that IIB matrix
model reproduces the standard perturbative series of superstring 
through $1/N$ expansion. Related model have been proposed in refs. [6-17]
and further
investigations on this model performed in refs. [18-26].
In this way, we have almost confirmed that IIB matrix model is a candidate of
nonperturbative definitions of superstring theory. Our next task is to understand
properties of the model better and to extract nonpertubative effects of
superstring by approximating the model properly or performing  numerical 
simulations.
In particular, we should clarify mechanism for dynamical generation
of space-time.
%To this end, we still need to clarify some properties of the model.
%In particular, it is significant to understand  how we should treat the zero mod%%the model should be defined strictly. We are interested in the cut-off %scale c%an be introduced in the model, which is required in taking the %double scaling l%
In this letter, as a first step to this end, we investigate $N_c=2$ case of 
IIB matrix model. We calculate the partition function and some correlation 
functions consisting of Wilson loops exactly.\footnote{A similar calculation of 
the partition function has been  performed in a different context 
in ref. \cite{GG}.
and given a different result. 
%Their result is different from ours.
}  We discuss some implications of our results.

IIB matrix model is defined by the action
\beq
S=-\frac{1}{g^2}(\frac{1}{4}Tr[A_{\mu},A_{\nu}]^2+\frac{1}{2}
Tr(\bar{\psi}\Gamma^{\mu}[A_{\mu},\psi])),
\label{action}
\eeq
where $A_{\mu}$ and $\psi$ are $N_c \times N_c$ hermitian matrices,
and $A_{\mu}$ and $\psi$ have ten-dimensional vector and Majorana-Weyl 
spinor index respectively.
We consider naively that we need to
restrict the eigenvalues of $A_{\mu}$'s in the path integration from
$-\pi/a$ to $\pi/a$, where $1/a$ is some infrared cut-off. 
%However, as we will %see below, any cut-off is unnecessary
%for the $N_c=2$ case.

In $N_c=2$, $A_{\mu}$ and $\psi$ can be expressed as\footnote{Here 
we do not treat the overall $U(1)$ part of matrices
proportional to $2 \times 2$ identity matrix, which gives the conservation
of total momentum in  correlation functions of Wilson loops.}
\beqa
A_{\mu}=\sum_{a=1}^{3} \frac{1}{2}\sigma^a A_{\mu}^a, \n
\psi=\sum_{a=1}^{3} \frac{1}{2}\sigma^a  \psi^a,
\eeqa
where $\sigma^a$'s are the Pauli matrices.
First let us consider the partition function for this case,
\beq
Z=\int \prod_{a=1}^{3} d^{10 }A^a d^{16} \psi^a \; e^{-S}.
\eeq
We can perform integrations over $\psi$'s and obtain the following
fermion determinant (pfaffian), up to some numerical constant.
\beqa
Pf&=&\frac{1}{g^{48}}
           (Tr([A_{\mu},A_{\nu}][A^{\nu},A_{\rho}][A^{\rho},A^{\mu}]))^4\n
   &=&\frac{1}{g^{48}}((A^1)^2 (A^2)^2 (A^3)^2-(A^1)^2(A^2 \cdot A^3)^2
                         -(A^2)^2(A^3   \cdot A^1)^2 \n
   &         &     -(A^3)^2(A^1 \cdot A^2)^2
                  +2 (A^1 \cdot A^2)(A^2 \cdot A^3)(A^3 \cdot A^1))^4.
\label{pfaffian}
\eeqa
To perform integrations over $A_{\mu}$'s, we set
\beqa
&&A_{0}^{3}=L \n
&&A_{i}^{3}=0 \; (i=1,\cdots, 9),
\eeqa
by using the Lorentz invariance. We also parametrize $A^1$ and $A^2$ as
\beqa
&&A^{1}_{0}=a_1, \;  A^{2}_{0}=a_2, \n
&&|A^{1}_{i}|=R_1, \; |A^{2}_{i}|=R_2, \;
\sum_{i=1}^{9} A^{1}_{i} A^{2}_{i}=R_1 R_2 \cos \theta.
\eeqa 
After the above prescription, the bosonic part of the action (\ref{action})
becomes
\beqa
S_{b}=\frac{1}{4g^2}(L^2(R_{1}^{2}+R_{2}^{2})
+R_{1}^{2} R_{2}^{2}\sin^2 \theta \n
\vspace*{3cm} +a_{1}^{2}  R_{2}^{2}+a_{2}^{2}  R_{1}^{2}
+2a_{1} a_{2} R_{1} R_{2} \cos \theta),
\eeqa
and the pfaffian (\ref{pfaffian})
\beq
\frac{1}{g^{48}} L^8 R_{1}^{8} R_{2}^{8} \sin^8 \theta.
\eeq
Here we have performed analytic continuations such as
\beq
L \rightarrow i L, \; a_1 \rightarrow i a_1 \; \mbox{and} \;
a_2 \rightarrow i a_2.
\eeq
The measure for the bosonic variables are evaluated  as
\beq
\int \prod_{a=1}^{3} d^{10 }A^a
=\int dL d\Omega_{3}^{(9)} L^9 \int da_1 da_2 \int dR_1 dR_2 d\Omega_{1}^{(8)} 
d\Omega_{2}^{(7)} d\theta  R_{1}^{8} R_{2}^{8} \sin^7 \theta.
\eeq

By gathering the above results and performing integrations over
$\Omega_1$, $\Omega_2$, $\Omega_3$, $a_1$, $a_2$ and $R_2$,
we obtain, up to an overall numerical constant,
\beq
Z=\frac{1}{g^{30}} \int dL L^{17} \int dR_1 d\theta
\frac{R_{1}^{15} \sin^{14}\theta}{(R_{1}^{2} \sin^2 \theta + L^2)^8}
\; e^{-\frac{1}{4 g^2} L^2 R_{1}^{2}}.
\eeq
Furthermore we can integrate out $\theta$ in the above expression. The result is
\beq
Z=\frac{1}{g^{30}}\int dL L^{16} \int dx \frac{x^7}{(x+L^2)^{\frac{15}{2}}}
                       \; e^{-\frac{1}{4 g^2} L^2 x},
\label{partitionfn}
\eeq
where $x=R_{1}^{2}$. After $x$ integration, the integrand of $L$
behaves as
\beq
L^9 L^{-24} \; \mbox{in the large $L$ limit,}
\label{largeLlimit}
\eeq
and as
\beq
L^{15} \; \mbox{in the small $L$ limit,}
\eeq
and has no singularities.  
Therefore the partition function is finite without introducing any cut-off. 
Indeed, the value of the expression (\ref{partitionfn})
is  $\frac{1}{g^{21}} 2^7 B(\frac{7}{2},4)\Gamma(\frac{9}{2})$.
We can also see that 
$\langle \sqrt{(A^3)^2} \rangle=\langle L \rangle \sim g^{1/4}$, 
where $g^{1/4}$ should be identified with $\alpha'^{1/2}$.
Note here that $L$ corresponds to the distance between two eigenvalues (diagonal
elements) of $A_{\mu}$'s. 
There are attractions in the long distance and repulsions in the short distance
between the two eigenvalues. The potential is logarithmic in these regions.
They constitute a
bound state, which represents "space-time" spreading over the Planck length.

By using a similar technique, we can calculate a correlation function consisting
of Wilson loops such as
\beq
\langle Tr(e^{i K^{\mu} A_{\mu}}) Tr(e^{-i K^{\nu} A_{\nu}}) \rangle,
\eeq
where $K^{\mu}$'s correspond to total momenta of strings.
The result is
\beq
\frac{1}{2} \left(1+\frac{W(|K|)}{W(0)} \right),
\label{correlationfn}
\eeq
where a function $W(x)$ is defined as
\beq
W(x)=\sum^{3}_{n=0} \left( \begin{array}{c} 3\\n \end{array} \right)
\Gamma \left(\frac{7}{2}+n \right) (-2)^{n}
\int_{0}^{\infty} dte^{-t^2-\frac{K^2}{16 t}} 
t^{m-3} H_{8-2n} \left( \frac{x}{\sqrt{2t}} \right)
\eeq
and $H_k(x)$ is the hermite polynomial.
Note that the result (\ref{correlationfn}) is also finite naturally
since Wilson loops
are bounded functions.   
We expect that it serves as a test for numerical simulations.

It is rather nontrivial that we obtain the finite partition function without
introducing any cut-off.
Let us consider two examples to clarify this point. The first one is the
reduced model of two-dimensional bosonic $U(N_c)$ Yang-Mills theory,
%\footnote{We
%are obliged to H. Kawai and J. Nishimura for many discussions on this model.}
which is soluble exactly and resembles supersymmetric theories as long as 
the counting of degrees of freedom goes. Its partition function is
\beq
Z_{2DYM}=\int dA^1 dA^2  e^{\frac{1}{2 g^2}Tr[A^1,A^2]^2}.
\eeq
We can diagonalize $A^1$ as
\beq
A^1=U \mbox{diag}(\lambda_{1},\lambda_{2}, \cdots, \lambda_{N_c})U^{\dagger},
\eeq
where $U$ is a unitary matrix.
Then we first perform the integration over the angular part $U$ and obtain
\beq
Z_{2DYM}=\int\prod_{i=1}^{N_c}d\lambda_{i}dA^2\prod_{i>j} (\lambda_i-\lambda_j)^2e^{-\frac{1}{g^2}\sum_{i>j} (\lambda_i-\lambda_j)^2 |A^{2}_{ij}|^2}.
\eeq
The integrations over the off-diagonal parts of $A^2$ lead to
\beq
Z_{2DYM}=\int\prod_{i=1}^{N_c}d\lambda_{i}dA^{2}_{ii}.
\eeq
Thus we find that in this case we indeed need a cut-off $1/a$ for the $2 N_c$ 
bosonic zero modes.

As the second example, we consider the one-loop effective action in the
diagonal background in IIB matrix model,
\beqa
A^{\mu}&=& \mbox{diag} (x^{\mu}_{1}, x^{\mu}_{2}, \cdots, x^{\mu}_{N_c}),\\
\psi&=&0.
\eeqa
We expand the action up to the second order in quantum fluctuations, 
which include only off-diagonal parts. It have been shown in ref. \cite{IKKT}
that in this case the one-loop determinant coming from bosons 
cancel against those from fermions and Faddeev-Popov ghosts. 
Therefore we obtain as the
partition function
\beq
Z=\int \prod_{i=1}^{N_c} d^{10}x_{i} d^{16}\psi_{ii}
\eeq
This gives zero due to $16 N_c$ fermionic zero modes. We expect naively
that in the more exact treatment the zeros from the fermionic zero modes
are cancelled by
infinities from bosonic zero modes and that
the partition function is nonzero.

We have shown explicitly that this is the case for $N_c=2$ 
and that the partition function is finite without any cut-off.
%Here let us consider for generic $N_c$ 
%integrations over diagonal parts of fermions after
%in the background.
%We can verify that for generic $N_c$ we obtain long-distance behavior
%such as $(\Delta x)^{-24(N_c-1)}$, where $\Delta x$ is a difference between
%two diagonal elements of bosonic matrices, when we integrate out diagonal parts
%of fermions after a one-loop calculation with the diagonal parts of fermions
%kept in the background.
%We can verify that by integrating out diagonal parts of fermions 
%after a one-loop
%calculation with the diagonal parts of fermions kept in the background,
%we obtain the long-distance behavior (\ref{largeLlimit}),
%which is derived from the exact result.
%This behavior is consistent with the long-distance behavior
%(\ref{largeLlimit}) for $N_c=2$, which is derived from the exact result.
We can verify that from one-loop calculation with the diagonal parts of
fermions treated more elaborately we obtain long-distance 
behavior suppressing infrared divergences.\footnote
{We would like to thank H. Kawai for discussing this point.} 
For $N_c=2$, that kind of one-loop calculation accounts for
the long-distance behavior
(\ref{largeLlimit}), which is derived from the exact result.
This suggests that long-distance behavior is controlled in general by that kind of calculation and there are no infrared divergences for generic $N_c$.
%and there are no infrared divergences for generic
$N_c$.
%since our
%exact result for $Nc=2$ suggests that 
%obtain similar long-distance behavior
%from integrations over diagonal
%parts of fermions
%after this one-loop calculation for this case.
%For generic
%$N_c$, the result of this calculation behaves as
%$(\Delta x)^{-24(N_c-1)}$, where $\Delta x$
%is a difference
%between two diagonal elements of bosonic matrices. Thus we hope that
%we do not need any infrared cut-off $1/a$ for an arbitrary $N_c$. 
%This is
%consistent with a possibility suggested by Kawai \cite{Kawai} that the gas of
%eigenvalues spreads depending on $N_c$ and that the string coupling constant is %determined automatically. 

In this article, we have shown that in $N_c=2$ case "space-time"
consisting of the two eignevalues indeed
exsits due to a proper potential between them without diverging nor collapsing.
%To understand the dynamical generation of space-time more deeply, finally, 
%we should clarify properties of
%many-body forces between the eigenvalues. To this end, we will treat 
%the larger $N_c$ cases, for the most simple example, the $N_c=3$ case.
To understand the dynamics of the eigenvalue gas
in the realistic large $N_c$ cases, we should clarify effects of entropy
and many-body interactions between
the eigenvalues. For the latter, it seems very useful to investigate the larger
$N_c$ cases intensively, for the simplest example, $N_c=3$.
We hope that we will report this subject elsewhere. 

%\vspace{2cm}
\newpage
One of us (A. T.) would like to thank H. Kawai for stimulating discussions
and comments.
He is also grateful to H. Aoki, S. Iso, Y. Kitazawa, J. Nishimura and T. Tada
for discussions. We are obliged to our colleagues at Osaka University,
especially H. Itoyama for discussions and reading the manuscript carefully.

\end{document}